\documentstyle[12pt,aaspp4]{article}

\def\be1#1{\hbox{${}^{1#1}$Be}}
\def\c1#1{\hbox{${}^{1#1}$C}}
\def\al2#1{\hbox{${}^{2#1}$Al}}
\def\cl3#1{\hbox{${}^{3#1}$Cl}}
\def\mg2#1{\hbox{${}^{2#1}$Mg}}
\def\k4#1{\hbox{${}^{4#1}$K}}
\def\ca4#1{\hbox{${}^{4#1}$Ca}}
\def\mn5#1{\hbox{${}^{5#1}$Mn}}
\def\fe#1{\hbox{${}^{#1}$Fe}}
\def\ni5#1{\hbox{${}^{5#1}$Ni}}
\def\i12#1{\hbox{${}^{12#1}$I}}
\def\sm14#1{\hbox{${}^{14#1}$Sm}}
\def\u23#1{\hbox{${}^{23#1}$U}}
\def\pu24#1{\hbox{${}^{24#1}$Pu}}

\def\scn#1#2{\hbox{$#1 \times 10^{#2}$}}

\def\order#1{\hbox{${\cal O} (#1)$}}
\def\mej{\hbox{$M_{\rm ej}$}}
\def\msw{\hbox{$M_{\rm sw}$}}
\def\mtot{\hbox{$M_{\rm tot}$}}
\def\msol{\hbox{$M_{\odot}$}}
\def\vej{\hbox{$v_{\rm ej}$}}
\def\pc{\hbox{${\rm pc}$}}
\def\avg#1{\langle #1 \rangle}
\def\muu{\hbox{$m_{\rm u}$}}

\def\beq{\begin{equation}}
\def\eeq{\end{equation}}
\def\beqar{\begin{eqnarray}}
\def\eeqar{\end{eqnarray}}

\def\la{\mathrel{\mathpalette\fun <}}
\def\ga{\mathrel{\mathpalette\fun >}}
\def\fun#1#2{\lower3.6pt\vbox{\baselineskip0pt\lineskip.9pt
  \ialign{$\mathsurround=0pt#1\hfil##\hfil$\crcr#2\crcr\sim\crcr}}}

\begin{document}

\title{GEOLOGICAL ISOTOPE ANOMALIES AS \\
SIGNATURES OF NEARBY SUPERNOVAE}

\author{John Ellis} \affil{Theoretical Physics Division,CERN \\
Geneva, Switzerland} 
\authoremail{JOHNE@cernvm.cern.ch}

\author{Brian D. Fields\altaffilmark{1}}
\affil{Department of Physics, University of Notre Dame \\
Notre Dame, IN 46556}
\altaffiltext{1}{also Institut d'Astrophysique de Paris, 98 bis Boulevard Arago
 \\
Paris 75014, France}
\authoremail{brian.fields@nd.edu}

\author{David N. Schramm\altaffilmark{2}}
\affil{University of Chicago, 5640 S. Ellis Avenue \\
Chicago, IL  60637}
\altaffiltext{2}{also NASA/Fermilab Astrophysics Center, FNAL Box 500 \\
Batavia, IL, 60510}
\authoremail{dns@oddjob.uchicago.edu}

\begin{abstract}
Nearby supernova explosions may cause geological isotope anomalies via
the direct deposition of debris or by cosmic-ray spallation in the
earth's atmosphere.  We estimate the mass of material deposited
terrestrially by these two mechanisms, showing the dependence on the
supernova distance.  A number of radioactive isotopes
are identified as possible diagnostic tools, such as \be10, \al26, \cl36,
\mn53, \fe{60}, and \ni59, as well as the longer-lived \i129, \sm146,
and \pu244.  We discuss whether the 35 and 60 kyr-old \be10 anomalies
observed in the Vostok antarctic ice cores could be due to supernova
explosions.  Combining our estimates for matter deposition with results of
recent nucleosynthesis yields, we calculate the expected signal from
nearby supernovae using ice cores back to \order{300} kyr ago, and we
discuss using deep ocean sediments back to several hundred Myr. In
particular, we examine the prospects for identifying isotope anomalies
due to the Geminga supernova explosion, and signatures of the
possibility that supernovae might have caused one or more biological
mass extinctions.
\end{abstract}


\section{Introduction}
The most violent events likely to have occurred in the solar
neighbourhood during geologic (and biological)
history could have been supernova
explosions.  The likelihood of such events has recently
been impressed upon us by the discovery that Geminga
is a nearby (Caraveo, Bignami, Mignani, \& Taff 1996)
and recent supernova remnant (Gehrels \& Chen 1993;
Halpern \& Holt 1992; Bignami \& Caraveo 1992; 
Bertsch et al.\ 1992).
If a supernova explosion occurred sufficiently close to Earth,
it could have dramatic effects on the biosphere
(Ruderman 1974).  Various
processes have been discussed, including an enhanced
flux of cosmic radiation and possible stripping of the Earth's
ozone layer followed by the penetration of solar
ultraviolet radiation (Reid, McAfee, \& Crutzen 1978;
Ellis \& Schramm 1995)
and absorption of visible sunlight by NO$_2$
(Crutzen \& Br\"uhl 1995),
which could be life-threatening,
and direct deposition of supernova debris.
Any attempt to identify supernova effects in one of the
many well-established mass extinctions must remain
speculation in the absence of tools to diagnose the explosion of a nearby
supernova using either the geophysical or the astrophysical
record.

This paper discusses isotope anomalies as possible geological
signatures of a nearby supernova explosion.
This is a not a new idea:  in fact it was the motivation for the
Alvarez search (Alvarez et al.\ 1980)
that discovered the iridium anomaly which
is now believed to be due to an
asteroid or comet impact (van den Bergh 1994)
at the time of the K-T transition that probably played a role in
the extinction that occurred then.  Moreover, \be10
isotope anomalies
corresponding to geological
ages of about 35 and 60 kyr before the present
have actually been discovered in
ice cores (Raisbeck et al.\ 1987; Beer et al.\ 1992;
Raisbeck et al.\ 1992),
and deep sea sediments
(McHargue, Damon, \& Donahue  1995; Cini Castagnoli et al.\ 1995).
Their interpretation        in terms of
one or more nearby supernova explosions has been
discussed
(Raisbeck et al.\ 1987;
Sonett, Morfill, \& Jokipii 1987;
Ammosov et al.\ 1991;
Sonett 1992;
Ramadurai 1993).
This paper is an attempt to update such studies  in
the light of the current understanding of supernova remnant evolution,
following supernova 1987A (reviewed in, e.g.,
Arnett et al.\ 1989; McCray 1993)
and the recent developments regarding
Geminga and the \be10 anomalies.

Whilst we consider here the general issues involved in detecting
any nearby supernova, we note that any event within about
$10$ pc would have had a profoundly deleterious effect upon
biology (Ellis \& Schramm 1995). Thus in our discussion we will
place special emphasis on the specific case of an event
at a distance of $\sim 10$ pc.

The total amount of material deposited by a nearby supernova
by both direct and indirect means is relatively small;
thus if one wants to avoid the large background of isotopes
produced during most of the Universe's history,
the most easily detectable isotopic signatures of such a supernova
are probably radioisotopes and their decay products.
A signature may appear as live and/or extinct radioactivity,
raising different issues for detectability.
In the case of live radiation,
the isotopes of interest must have lifetimes less than about
$10^9$ yr,  if one
is interested in events that could have had a significant effect
on the Earth's biosphere.
If, in addition, one is interested in a correlation with one of the
well-documented mass extinctions, the isotope lifetime should be
longer than about $10^7$ yr in order for it still to be present.
Shorter-lived extinct radioactivities, it turns out, are
unlikely to be detectable.

The possible candidate isotopes with long lifetimes include
\i129, \sm146, and \pu244.  If one is interested
in understanding the origin of the Vostok \be10 anomaly, the lower
limit on the lifetime may be reduced to about $10^4$ yr, in which case
\al26, \cl36, \ca41, \mn53, \fe{60}, and \ni59
may be added to the list of relevant isotopes.
In \S \ref{sec:sig},
we calculate explicitly the supernova signatures,
as well as the background, for
all isotopes expected to be observable in the Vostok ice cores.

There are two ways in which a nearby supernova explosion could produce
anomalous isotopes: either indirectly as cosmic ray spallation
products, which would be more important for light isotopes such as
\be10, or directly via the deposition of supernova debris, which would
be more important for intermediate-mass isotopes such as \ca41 and
\fe{60}.  The very heavy r-process isotopes are probably associated
with supernovae (Meyer et al.\ 1992), but alternative sources are also
possible (Meyer \& Schramm 1986).  Thus discovery of r-process
anomalies that correlated with an intermediate-mass anomaly would help
to establish supernovae as the astrophysical r-process source.  The
relative importance of these classes of anomalies depends on the
distance at which the supernova exploded, since supernova ejecta are
slowed down and eventually stopped by the ambient pressure
of the interstellar medium
(ISM). Later in this paper, we give a quantitative discussion of the
ratio of spallogenic and direct deposition isotopes as a measure of
the distance of any putative supernova explosion.

We then discuss the usefulness of this diagnostic tool for understanding
the origin of the geologic \be10 anomalies,
and review the prospects for extending anomaly searches back
to \order{300} kyr ago using older ice cores,
and back to \order{500} Myr ago using deep ocean sediments.
In particular, we discuss whether the \be10 anomalies
could be associated with the supernova explosion that
created Geminga (Halpern \& Holt 1992; Bignami \& Caraveo 1992; 
Bertsch et al.\ 1992;
Gehrels \& Chen 1993).
This seems unlikely, in view of the spin-down age of
Geminga and the size of the local bubble in the interstellar medium,
but cannot be excluded in view of the
large uncertainties in the Geminga age estimates.
Moreover, this possibility can be explored
by looking for correlated anomalies as discussed in
\S \ref{sec:isotopes}.  Even in the absence of
such a correlation with the \be10 anomalies, this technique
could be used to search for a geological signature of the Geminga
explosion if it occurred  up to \order{300} kyr ago, as generally believed.

\section{Isotope Production}

Nearby supernovae can introduce radioisotopes to the earth
by two mechanisms:  direct deposition of material from
the ejected shell, or spallative production in the
earth's atmosphere (i.e., cosmogenic production)
due to the supernova's enhancement of the local cosmic ray flux.
In this section, we discuss the physics of both mechanisms and estimate
the total mass deposited on the earth.
We will then use these results in \S\ref{sec:sig} to determine
experimental signatures of these mechanisms in terms of
ice-core and deep-ocean sediment observables.

\subsection{Direct Deposition:  Supernova Remnant Dynamics}

Consider the direct terrestrial deposition of
the supernova blast matter.  Note that this in fact contains two components:
(1) material ejected from the supernova itself, and
(2) material swept up by the ejecta as it traverses the
ISM on its way to Earth.
Imagine a supernova exploding at a distance $D$ from Earth and
ejecting a mass \mej\ of which a fraction $X_i^{\rm SN}$ is composed of
isotope $i$.  If the amount of matter swept up
is \msw, with composition (mass fractions)
$X_i^{\rm ISM}$, then the total
mass arriving in the shell as it reaches the Earth is
$\mtot = \mej + \msw$.  The composition
of this material is a weighted average of the abundances in
each component:
$X_i = (X_i^{\rm SN}\mej+X_i^{\rm ISM}\msw)/\mtot$.
The proportion of this matter that reaches the Earth
is just given by the fraction of the solid angle the Earth subtends.
The mass in $i$ deposited terrestrially is thus
\beqar
\label{eq:mdep}
\nonumber
M_i^{\rm dep} & = & f_{\rm dep} \ X_i \
\left(\frac{R_{\oplus}}{2D}\right)^2 \mtot \\
\nonumber
  & = & 2.3 \times 10^{13} \ {\rm g} \ \, f_{\rm dep} \ X_i \\
  & & \times
        \left(\frac{D}{10 \ \pc}\right)^{-2} \
        \left(\frac{M_{\rm tot}}{100 \; \msol}\right)
\eeqar
Note that the deposited mass $M_{\rm dep}$ depends on the
distance $D$ to the supernova via the contribution of the swept
material \msw; this dependence can be understood in terms of
supernova remnant evolution, and will be considered shortly.
Note also that we have inserted in eq.\ (\ref{eq:mdep}) a factor
$f_{\rm dep} \le 1$ to account for partial exclusion of ejecta
{}from the solar cavity due to
the solar wind.

Equation (\ref{eq:mdep}) shows that the order of magnitude of the
total mass deposited is $10^{13}$ g, or about 100 million tons.
This is quite
small compared, for example, to the K-T object's estimated
mass of $2.5 \times 10^{17}$ g (van den Bergh 1994).
Thus one cannot hope to find evidence for this deposited matter
using the techniques of Alvarez et al.\ (1980),
which involve searches for isotopic anomalies in
stable nuclei.  In our case, the amount of material deposited is too
small for such anomalies to be detectable above the background
material with terrestrial composition.  Thus we are instead
driven to look for isotopes for which the background is very low,
namely those which are unstable but long-lived:  the radioisotopes.
Below (\S \ref{sec:sig}), we will consider in detail
both live and extinct radioactivities.  For the moment, one need only
keep in mind that the species of interest are unstable, and thus
it remains to be seen which ones have the best production abundances,
the lowest backgrounds, and the best lifetimes to be useful
diagnostics of nearby supernovae.

The propagation of the shock is indicated in eq.\ (\ref{eq:mdep})
via the implicit dependence of \mtot\ on $D$; in fact we can be
more specific about the shock's mass and motion.
The motion of real shocks, and their interaction
with the ISM, is complicated; recent detailed
discussion can be found in, e.g., McKee (1988),
and Chevalier \& Liang (1989).
The propagation phases include:  free expansion for $\sim 4$ pc
until the ejecta has swept up about its own mass -- subsequent to this
the ISM dominates the mass and composition; then
adiabatic (Sedov) expansion until radiative losses become
important, and finally the momentum-conserving ``snow plow'' phase.
In fact, we will not even need to delve into the details of these phases.
We only wish to estimate
the swept up mass \msw, and in
{\it all} of these phases the ISM is swept up by the shock.
For the purposes of making order of magnitude estimates we
construct a simplified model as follows .

The total mass
ejected or swept up
at distance $D$ from the supernova is
\beqar
\label{eq:mtot}
\nonumber
\mtot & = & \mej + \msw \\
  & = & \mej + \frac{4\pi}{3} \; \rho_{\rm ISM} \; D^3 \ \ .
\eeqar
To determine the swept-up mass, choosing an appropriate value
for  $\rho_{\rm ISM}$ (or equivalently $n_{\rm ISM}$)
is essential.  Unfortunately, there is a wide range of
reasonable choices.  The average ISM number density is
$\sim 1 \ {\rm cm}^{-3}$, but within hot, supernova-induced
bubbles, the density is closer to $\sim 10^{-3} \ {\rm cm}^{-3}$.
And while the solar system is presently located on the edge of
such a bubble (Frisch 1994), it may have only arrived there
recently, and at has probably traversed may
different environments on the timescales of hundreds of million
years associated with mass extinctions.  Nevertheless, we conservatively
adopt the lower value as a fiducial one; in fact, we will see that
this choice impacts only the long-lived, supernova-produced
radioisotopes.

The accumulation of mass continues until the end of the  snow-plow
phase when the shock finally stops; we wish to estimate the distance
at which this occurs.  To do so, we note that in this phase
the shock slowing is determined by momentum conservation.
Let us assume that the transition to this
 phase from the adiabatic expansion phase happens
at a distance $D_{0} \simeq 20$ pc, with velocity
$v_{0} \simeq 100$\  km/s, mass
$M_{0} \simeq 4\pi/3 \; \rho_{\rm ISM} D_{0}^3 \simeq 1000 \msol$,
and time $t_{0} \sim 40$ kyr (as given, e.g., in Spitzer 1978).
The transition momentum is thus $M_0 v_0$, and setting
this equal to $\mtot v$ we have
\beq
\label{eq:v}
\mtot = \frac{v_0}{v} \, M_0 \ \ .
\eeq
This accretion process continues until the shock pressure drops
to a level
comparable to that of the ISM, at which point the shock stops.
An estimate of the distance scale for the shock quenching
gives a final radius $D_{\rm f} \simeq 70$ pc for an
ISM temperature of $10^4$ K.

Even if the shock is stopped in the ISM
due to ISM thermal pressure, the solar system may pass through it.
But in this case the material
will be repelled by the solar wind, which at 1 AU has a much
higher pressure. It is also possible that the shock may be
repelled by the solar wind even before it is stopped by the ISM{}.
In eq.\ (\ref{eq:mdep}) we have indicated
this exclusion from the earth by the factor $f_{\rm dep}$,
but we may approximate its effect
by simply finding a
smaller $D_{\rm max} < D_{\rm f}$ appropriate for
the solar wind pressure (i.e.,
we will put $f_{\rm dep} = 1$ for $D \le D_{\rm max}$ and
$f_{\rm dep}=0$ otherwise).

We thus need to estimate $D_{\rm max}$.
Equating the ejecta pressure
$P_{\rm ej} \sim \mtot v/(D^2 \Delta t) = \mej \vej/(D^2 \Delta t)$
with the solar wind pressure
$P_{\rm sw} \sim \muu v_{\rm sw} \Phi_{\rm sw}$ gives
a maximum range of $\sim 16$ pc.
Note, however, that this calculation
assumes the worst-case geometry, namely that the shock encounters
the wind perpendicularly on its way to the earth.  A more oblique
angle allows more penetration and so a higher $D_{\rm max}$.   This
effect will be important even if the explosion happens in the plane
of the ecliptic as long as the shock duration $\Delta t > 1$ yr, allowing
the earth to encounter regions at these oblique angles.  Furthermore,
one generically expects the explosion to be be out of the ecliptic.
A detailed analysis of the possible geometries is beyond the scope of
this paper, but it is clear it will lead to a larger $D_{\rm max}$
that this simple estimate.  To allow for this and to recognize the
uncertainties of the calculation, we will relax the limit by a factor
of 3 for the purposes of discussion, and so we have
\beq
D_{\rm max} \simeq 50 \ {\rm pc} \
   \left( \frac{\mej}{10 \msol} \right)
   \left( \frac{\vej}{v_{\rm sw}} \right)^{1/2}
   \left( \frac{\Delta t}{1 kyr} \right)^{-1}
\eeq
where we have used $\Phi_{\rm sw} = 3 \times 10^8 \ {\rm protons} \;
{\rm cm}^{-2} \; {\rm s}^{-1}$ and $v_{\rm sw} = 400 \ {\rm km/s}$.

Since sweep-up is effective until
the shock dies, we will model the spatial dependence of the
deposited material by
using eq.\ (\ref{eq:mtot}) until the distance $D_{\rm max}$, which
we will take
to be a sharp cutoff.
Beyond $D_{\rm max}$, the only material deposited
is of cosmogenic origin, which we will see in the next section is
a much smaller amount.  Thus the cutoff sets a crucial
distance scale, above which the signal becomes
very much weaker.  A plot of this behaviour appears in figure \ref{fig:mvd}.

Figure \ref{fig:mvd} points up a
striking feature of the direct deposition mechanism
for the case of an explosion within a dense ISM.
In the regime $10 \ {\rm pc} \la D \la D_{\rm max}$,
the total shock mass varies as
$\mtot \approx \msw \sim D^3$,
while the Earth's solid angle
with respect to the supernova goes as $D^{-2}$.
Consequently,
the deposited mass actually {\it increases}\ linearly with $D$
for the larger distances.  On the other hand, the deposition of cosmic
and $\gamma$ radiation monotonically decreases.  Since the latter
are the cause of the supernova's biohazard, then at the distance of
$\sim 10$ pc most interesting for mass extinctions, the direct deposit
material is in fact near its minimum amount.  To be sure, the variation
is relatively small, namely less than an order of magnitude.
Nevertheless, it is ironic that some relatively harmless supernovae
could in fact leave larger signals than a catastrophic nearby event.

\subsection{Direct Deposition:  Composition}

Note also that swept-up material has the composition of the ISM,
which is very different from that of the supernova ejecta.
Further, the ratio of these two sources depends on the
amount of material swept up, and thus on the distance
to the supernova.  Specifically, the ratio is
\beqar
\nonumber
\frac{M_i^{\rm sw}}{M_i^{\rm ej}} & = &
    \frac{X_i^{\rm ISM}}{X_i^{\rm SN}} \
    \frac{4\pi/3 \; D^3 \ \rho_{\rm ISM}}{\mej} \\
  & = &  10.3 \ \frac{X_i^{\rm ISM}}{X_i^{\rm SN}} \
    \left(\frac{D}{10 \ {\rm pc}}\right)^3 \
    \left(\frac{n_{\rm ISM}}{1 \ {\rm cm}^{-3}}\right) \
\eeqar
i.e., the swept-up component increases like $D^3$ relative to the
supernova ejecta, and is dominant
before a distance of 10 pc
{\it if there is significant abundance of
$i$ in the ISM}.

In fact, one should distinguish three categories of directly
deposited (radio)isotopes,
according to
their production sources and lifetimes.
First, there are the isotopes which
are not significantly produced by supernovae, but are created by
cosmic-ray interactions, e.g., \be10.
(We will treat cosmogenic production separately in \S \ref{sec:cgen}.)
As these isotopes are absent from the ejecta,
we are only concerned with the nuclei accumulated by
the sweeping of material
in ISM.  These will have an equilibrium abundance
in the ISM,
given by the cosmic ray production rate
averaged over the lifetime.  It turns out, however,
that the lifetimes are
short enough that this component is negligible compared to the
{\it in situ} cosmogenic component (which also benefits from
the target abundances being atmospheric and so dominated
by N and O, which are the main spallative \be10 progenitors).

Next we consider radioisotopes that {\it are} produced by supernovae.
These fall into two classes depending on the lifetime.  Long-lived
isotopes will have a significant ISM abundance, as the products of
many supernovae will accumulate during a lifetime; thus
these will appear in the swept matter which will be the dominant source
for long-lived isotope deposition.  Shorter-lived
isotopes, on the other hand, will die out too soon to have a large
ISM abundance, and so the deposition will be dominated by the supernova
ejecta.

The separation of these categories can be seen by computing the
swept contribution to supernova radioisotopes.  This is quite similar to
the swept spallogenic nuclide calculation.  The ISM equilibrium
density of a supernova isotope $i$ is
\beq
\rho_i = \lambda \tau_i \ \frac{X_i^{\rm SN} \mej}{V_{\rm gal}}
\eeq
where $\lambda \simeq (100 \ {\rm yr})^{-1}$  is
the galactic supernova rate, and $V_{\rm gal} = \pi R_{\rm gal}^2 h$
is the volume of the galactic disk with radius $R_{\rm gal} \simeq$ 10 kpc
and scale height $h \simeq$ 100 pc.
The total swept-up mass of $i$ is $M_i^{\rm sw} = 4\pi/3 \; \rho_i D^3$,
and the ratio of the swept to ejected mass in $i$ is
\beqar
\label{eq:longt}
\nonumber
\frac{M_i^{\rm sw}}{M_i^{\rm ej}}
  & = & \frac{4}{3} \lambda \tau_i \ \frac{ D^3}{R_{\rm gal}^2 h} \\
  & = &   1.3 \times 10^{-4}  \
         \left(\frac{\tau_i}{1 \ {\rm Myr}}\right) \
         \left(\frac{D}{10 \ {\rm pc}}\right)^3 \
\eeqar
which is small for moderate lifetimes; thus for isotopes having
$\tau_i \la$ Gyr, the ejecta composition dominates.  However, if
$\tau_i \ga 1$ Gyr and $D \ga 20$ pc, the swept component dominates
{\it if the explosion does not
occur within a rarefied bubble}.
These very long-lived isotopes are the best signatures of very
ancient mass extinctions; thus it is fortuitous that for just these
nuclides there can be a significant addition to their supernova ejecta
abundance.

Note that the different classes of isotopes have different distance
dependences.  In particular, those which are dominated by the ejecta
drop off as $D^{-2}$, while those dominated by swept matter
increase like $D$.  Thus measurements of
each of these types provides a independent
way of determining the
distance to the supernova; moreover, their ratio
provides an important consistency check.
Indeed, it is conceivable that the problem could be turned around
and one could learn about supernova ejecta by comparing ratios
of sedimentary radionuclides.

\subsection{Cosmogenic Production}
\label{sec:cgen}

The directly-deposited material
is to be compared to cosmogenic production from the enhanced cosmic rays
coming from the supernova.
An exploding supernova will invest some fraction $\epsilon \simeq 0.01$ of its
mechanical energy
in the production of cosmic rays; we will put
\beq
U_{\rm CR} = \epsilon \ U_{\rm SN}
\eeq
where $U_{\rm SN}$ is the kinetic energy of the blast wave.
If the average cosmic
ray kinetic energy is $\avg{E}_{\rm CR} \simeq 1$ GeV, then
the total cosmic ray exposure at Earth (i.e., the time-integrated
flux, or the fluence) is just
\beq
\Phi \;  \Delta t = \xi_{\rm CR} f_{\rm CR} \
\frac{U_{\rm CR}/\avg{E}_{\rm CR}}{4\pi D^2}
\eeq
where $\xi_{\rm CR} \le 1$ accounts for
losses due to propagation to the solar system, and $f_{\rm CR}$,
in analogy to $f_{\rm dep}$, allows
for exclusion from the solar cavity.

Note that the propagation is
very different from that of the blast material:
since the cosmic rays are much more diffuse and have a lower pressure,
they do not sweep up matter but move through it diffusively, spiraling around
local magnetic field lines.  The most significant means of
cosmic rays losses in transit are
due to ionization losses to the ISM; however, for the
pathlengths important here, these are completely negligible.
Thus cosmic ray losses in transit are minimal,
so we will put $\xi_{\rm CR}=1$ henceforth.
The physics behind $f_{\rm CR}$ is an accounting of the
solar wind's exclusion of cosmic rays; this is of course
the well-known solar modulation first described by
Parker (1958), and more recently re-examined by
Perko (1987).  For the purposes of estimation, we
will simply note that the integrated flux decreases by
roughly a factor of 10 from its interstellar value.
Thus we take $f_{\rm CR} = 1/10$.

The total number of cosmic-ray interactions with the Earth is
$\Phi \Delta t \; 4 \pi R_{\oplus}^2$; the fraction of these that
produce isotope $i$ in the process $j+k \rightarrow i$
is given by the branching ratio
$y_j^{\rm CR} y_k^{\rm atm} \sigma_{jk}^i/\sigma_{tot}$,
the ratio of spallogenic production of $i$ to the total inelastic
cross section
multiplied by the cosmic-ray and atmospheric abundances $y_j^{\rm CR}$
and $y_k^{\rm atm}$,
respectively.
It will be useful to introduce the definition
\beq
Y_i =  \sum_{jk} y_j^{\rm CR} y_k^{\rm atm}
   \frac{\sigma_{jk}^i}{\sigma_{\rm tot}}
\eeq
which amounts to a weighted branching ratio
for spallation production of $i$,
summed over all production channels;
a tabulation of $Y_i$ for many isotopes
of interest is found in O'Brien et al.\ (1991).
Then cosmic rays from a nearby supernova will
have a mass yield of isotope $i$ of
\beq
\label{eq:cosgen}
M_i^{\rm CR} = f_{\rm CR} \ A_i \ Y_i \
  \left(\frac{R_{\oplus}}{D}\right)^2 \mej \
  \left(\frac{\epsilon U_{\rm SN}}{c}\right)^2 \ \ .
\eeq

It is of interest to compare the strength of the two mechanisms,
direct deposition versus cosmogenic production.
As we have already noted, if the species is not produced
in supernovae, the cosmogenic component dominates
the contribution from swept ISM material.  However,
when there {\it is} a significant supernova contribution,
a straightforward analysis shows that this component dominates
that due cosmic rays; this reflects the fact that cosmic ray
spallation is a very inefficient mechanism for nucleosynthesis,
and is only relevant for nuclides which have no other
known astrophysical sources (e.g., $^6$Li, Be, and B).

The signatures of the different mechanisms may also be staggered in time.
In both cases the terrestrial signal will be delayed
after the supernova explosion
by the propagation time of the shock and of the cosmic rays.
This time of flight is significant for the shock,
which can take up to 100 kyr to go 20 pc.  On the
other hand, in the simplest picture the cosmic rays 
propagate diffusively ahead of the
shock.  They are much more rapid, traversing 20 pc in $\sim 1$ kyr.
Thus one does not expect the direct deposition and cosmogenic 
signals to be coeval.  However, if the cosmic rays, in the process
of acceleration, remain concentrated in the vicinity of the shock,
then there may sill be a cosmogenic signal simultaneous to
the direct deposition.

\section{Signatures and Their Detectability}
\label{sec:sig}

When some amount of a radioisotope is deposited in the Earth's
atmosphere, it will eventually precipitate out and accumulate in the
ice cores and the sea sediments.  Analysis of this material counts
the number of atoms, or rate of decays, per gram of ice or sediment.
In this section,
we estimate
the magnitude of the expected signal from a nearby supernova.

In the following, we will assume that the material
deposited in the atmosphere
will precipitate out uniformly around the Earth's surface.  This ignores
important considerations of the details of the mixing of atmosphere
and any chemical fractionation taking place during its deposition.
These effects can be important ones, as noted by, e.g.,
Beer, Raisbeck, and Yiou (1991)
in their discussion of \be10.  Despite these difficulties,
we forge ahead to see what sort of signature we would naively expect.
Clearly, however, a detailed treatment must address the issue
of chemical, atmospheric, geophysical and even biological effects.

\subsection{Live Radioactivity}
Thus far we have computed the total mass deposited at the earth by
a nearby supernova via the relevant mechanisms.
What is actually measured is the number of atoms, or of
decays, per gram of sediment.  Before making the connection
between the deposited mass and its final sedimentary abundance,
a word is in order about the experimental options and their
sensitivities.  A typical
sensitivity for mass spectrometry measurements of the
number of rare atoms per gram of bulk
material (call it $\Lambda$)
is around $\Lambda_{\rm min} \sim 10^4$ atoms/g.
Of course, the determination of $\Lambda$ is necessarily destructive.
On the other hand, one can perform a non-destructive measurement
of radioisotopes by measuring the decay rate (i.e., the activity).
The relation
between the two is simply
\beq
\label{eq:dpm}
\Gamma_i = \Lambda_i/\tau_i \ \ ,
\eeq
with
$\Gamma$ the decay rate per gram of bulk material.  Typical
sensitivities are $\Gamma_{\rm min} \sim 10$ dpm/kg
(dpm = decays per minute).
For a lifetime of 1 Myr, this threshold corresponds to
an effective number count threshold of 
$2 \times 10^9 {\rm atoms/g} \sim 10^5 \Lambda_{\rm min}$.  It is clear
that the mass spectrometry
techniques for  counting rare atoms are much more
favorable for our purposes.  Thus we suggest this method,
unless destructive tests are unavailable or unreliable.

We now wish to connect our calculation of total mass deposition with
the observables.
If a mass $M_i$ of isotope $i$ is deposited on the Earth, on average
it will precipitate out with a surface density $M_i/4 \pi R_{\oplus}^2$.
This will happen over the time $\Delta t$ it takes for the Earth
to receive the material, either directly as the supernova blast passes
through, or indirectly as the cosmic rays arrive.
If the bulk of the sediment or ice accumulates with a density
$\rho$ and its height increases at a rate $dh/dt$, then over a time
$\Delta t$ the surface density of the new sedimentation is
$\rho \; dh/dt \; \Delta t$.  Thus the number
of supernova radioisotopes per unit mass of terrestrial sedimentation
is
\beq
\label{eq:npm}
\Lambda_i = \frac{1}{A_i} \
\frac{M_i/\muu}{4\pi \rho \; R_{\oplus}^2 \; dh/dt \; \Delta t}
\eeq
where $M_i$ will depend on the deposition method, as we now discuss.

For short-lived direct-deposition isotopes
produced by supernovae, we have
$M_i = X_i^{\rm SN} \mej$, and so
\beqar
\label{eq:ddsig}
\nonumber
\Lambda_i & = &  5.0 \times 10^{7} \ {\rm atoms \ g}^{-1} \\
& & \times
       \left(\frac{X_i^{\rm SN}}{10^{-5}}\right) \;
       \left(\frac{\Delta t}{1 \ {\rm kyr}}\right)^{-1} \;
       \left(\frac{D}{10 \ {\rm pc}}\right)^{-2}
\eeqar
for $A_i = 50$ and $D \le D_{\rm max}$.
In eq.\ (\ref{eq:ddsig}) we have assumed a sedimentation density
$\rho = \rho_{\rm ice} \simeq 1 \ {\rm g \ cm}^{-3}$
and rate $dh/dt = 1$ cm/yr, in accordance with
the Raisbeck et al.\ (1987) Vostok measurements.
We see that the signature
is far above threshold, indicating that there should be a
strong signal, though not necessarily via decays. In the case
of \al26 in ice cores, we find
$\Lambda_{26}^{\rm ice} \simeq 10^{8} \ {\rm atom} \ {\rm g}^{-1}$
at $D = 10$ pc, which
is very much larger than the ice core \be10 spike amplitude.
Thus we predict that, if the ice core events were nearby supernovae within
direct-deposition range, the signal in \al26 and other
supernova-produced radioisotopes should be observable.

For the cosmogenic component produced {\it in situ}, we have
\beqar
\nonumber
\Lambda_i & = & f_{\rm CR} \epsilon \
  \left(\frac{U_{\rm SN}}{\muu c^2}\right)  \
  \frac{Y_i}{4 \pi \rho D^2 dh/dt \ \Delta t} \\
\nonumber
 & = &  7.7 \times 10^6 \ {\rm atoms} \ {\rm g}^{-1} \\
 & &  \times
       \left(\frac{Y_i}{10^{-2}}\right)
       \left(\frac{\Delta t}{1 \ {\rm kyr}}\right)^{-1}
       \left(\frac{D}{10 \ {\rm pc}}\right)^{-2}
\eeqar
using a value of $Y_i$ appropriate for \be10 in ice cores.

A similar approach can be used to estimate the possible
isotope signal in deep-ocean sediments, which precipitate
at a rate $dh/dt$ typically $10^{-3}$ of the rate of
accumulation of ice cores,
and may provide a fossil
isotope record extending back several hundred Myr.
The longer time scale means that one should concentrate on
longer-lived isotopes, to avoid a strong suppression
of the decay rate by an overall decay factor $e^{-(t_0-t_d)/\tau_i}$,
where $t_0$ ($t_d$) is the time at present
(at deposition)\footnote{The optimal choice is different for
the decay rate, which has
$\Gamma \propto e^{-(t_0-t_d)/\tau_i}/\tau_i$ and so is
maximized by $\tau_i = t_0-t_d$.  In practice this makes little difference
given the paucity of available radionuclei with $\tau \ga 10^8$ yr.}.
{}From this point of view, the optimal isotope lifetime
should be as long as possible, with an upper limit of about
the age of the earth (to assure that any initial
protosolar abundance has decayed away).
A catalog and discussion of isotope candidates can be found
in \S\ref{sec:isotopes}.

For ocean sediments, there is a lower limit
to the time resolution $\Delta t \ge 1$ kyr, the origin of which
is biological.  Namely, as noted in Beer et al.\ (1991)
small organisms dig into the sea floor and disturb it for
depths of a few cm, corresponding to a time of $\sim$ kyr.
This effect, known as  ``bioturbation,'' is an
example of the possible subtleties that must be addressed
in a more detailed account of our subject.  This
particular effect is presumably not
a problem with ice core samples, though they have their
own environmental peculiarities.

We thus re-emphasize that the above discussion does not
take into account possible fractionation due to chemical,
atmospheric, geophysical or even biological effects. Given
the longer time scales and greater exposure to such effects, the
assumptions of uniform deposition and stratification made above
are more questionable than for ice cores, and our estimate
eqs.\ (\ref{eq:npm},\ref{eq:dpm})
could be depleted by such effects. However, the possibility of
fractionation also suggests that the isotope abundances could
even be enhanced in some favourable cases. A detailed study
of the likelihoods of fractionation for the above-mentioned
isotopes goes beyond the scope of this paper.

In figure \ref{fig:obs} we plot the expected signal for
both kinds of deposition as a function of supernova distance.
Also indicated is a rough estimate of the experimental sensitivity,
as well as a calculation of the background cosmogenic production
due to galactic cosmic rays (discussed below in \S\ref{sec:crbg}).

\subsection{Extinct Radioactivity}

The technique here is similar to the one used by the Alvarez search
(Alvarez et al.\ 1980).  Consider a parent isotope $^iP$ (e.g., \al26) which
decays to a daughter isotope $^iD$ (e.g., \mg26).
A signal of the presence of $^iP$ would be a correlation of a
$^iD$ excess with the $P$ elemental abundance,
both measured in a ratio to the major isotope of $D$ (e.g., Mg).
E.g., one finds \mg26/\mg24 to be positively correlated with
Al/Mg; this allows one to deduce the protosolar \al26 abundance
(Lee, Papanastassiou, \& Wasserburg 1977).

For this procedure to work, the variations $\delta ^iD/D$ in the daughter
isotopic fraction must be detectable and not due to fractionation;
i.e., the {\em variations} must be at least of order of a percent.  This
means that the SN contribution to  $^iD$ must be at least of order
$^iD_{\rm SN} \ga 0.01 ^iD_{\rm BG}$; expressed in terms of number compared to
Si, we  have
$^iD_{\rm SN}/{\rm Si} \ga 0.01 (^iD/D)(D/{\rm Si})_{\rm BG}$.
If we take typical abundances of $D/{\rm Si} \sim 10^{-2}$,
and $^iD/D \sim 0.01$, we get a limit of $^iD/{\rm Si}_{\rm SN} \ga 10^{-4}$.
But in sediments we have signals of order $\Lambda_i \sim 10^9$ atom/g,
i.e., ${}^{i}D$ atoms are extremely rare.
Hence, even if the
sediment is only 1\% Si, this means an abundance of
$^iD/{\rm Si} \le 10^{-11}$, which is much less than the minimum
detectability.  So it appears that extinct radioisotopes
will have too feeble a signal to be measurable.

\section{Background Sources}

Radioisotope backgrounds arise from two mechanisms:  the normal
cosmogenic production in the atmosphere, as well as terrestrial
radiological production
due to fission of ambient heavy nuclei such as uranium.

\subsection{Cosmic Ray Background}
\label{sec:crbg}

Any signature we find must lie above the background of
radioisotopes continually produced in the atmosphere
by normal galactic cosmic rays, which is
just the usual cosmogenic production.
This problem has been well-studied and is summarized in,
e.g., O'Brien et al.\ (1991).  For our purpose, we may use the machinery of
the previous two sections to derive that the
rate of background atmospheric production of isotope
$i$ is
\beq
\frac{d}{dt}N_i^{\rm BG}  =  4\pi Y_i \; R_{\oplus}^2 \; \Phi_{\rm p}
\ \ ,
\eeq
using the notation of \S \ref{sec:cgen}, and where $\Phi_{\rm p}$ is
the total (modulated) cosmic-ray proton flux.
If this is incorporated into sedimentation or ice
with a surface density accumulating at a rate $\rho dh/dt$,
then the number of atoms per unit mass is
\beq
\label{eq:crbg}
\Lambda_i^{\rm BG} =  Y_i \ \frac{\Phi_{\rm p}}{\rho dh/dt} \
\eeq

We can check the calculation by
estimating the background production of \be10, for which
$Y = 1.36 \times 10^{-2}$.
We take a cosmic-ray proton flux of
$\Phi_{\rm p} = 10 \ {\rm cm}^{-2} \ {\rm s}^{-1}$.
With an ice density of $1 \ {\rm g} \ {\rm cm}^{-3}$ and
a deposition rate of $1 \ {\rm cm} \ {\rm yr}^{-1}$, we have
\beq
\Lambda_{\rm Be}^{\rm BG} \simeq 4 \times 10^6 \
{\rm atoms} \ {\rm g}^{-1}
\eeq
in rough agreement with the \be10 concentrations measured in the
ice cores. The fact that this simple estimate is apparently
too high by a factor of about two could be due to the
geomagnetic cutoff on some cosmic rays at low latitudes,
so that the average flux over the Earth's surface is reduced.
Such a possible error is smaller than other uncertainties in
our estimates; we will account for it here by lowering the
effective cosmic ray flux to $5 \ {\rm cm}^{-2} \ {\rm s}^{-1}$.

One may also estimate the \al26 background by this method;
O'Brien et al.\ (1991) calculate a cosmogenic production ratio
of $\al26/\be10 \simeq 2 \times 10^{-3}$.
This gives a cosmogenic \al26 background of order 300 atoms/g.
However, while this background is lower, the expected supernova signal
would be stronger than that of \be10 (if \al26 is produced in supernovae).
Thus, an \al26 signature would be very large and would far exceed background.

To make this point more broadly, we
now compare the
background due to galactic cosmic rays to the signals
of supernova deposition mechanisms.
For direct deposition of pure supernova products,
a straightforward comparison of
equations (\ref{eq:npm}) and (\ref{eq:crbg})
shows the signal to be very much larger
than the background.  To wit, for all cases of
interest, we estimate the signal-to-background ratio
to be $\ga 10^5$.  Direct deposition should thus be readily observable
if it has occurred.

For cosmogenic production,
the signal-to-background ratio is simply the efficiency
for the supernova to produce cosmic rays times the ratio
of the supernova cosmic ray flux to the galactic cosmic ray
flux.  Specifically,
\beqar
\label{eq:cgsb}
\nonumber
\frac{\Lambda_i^{\rm CR}}{\Lambda_i^{\rm BG}}
 & \simeq & f_{\rm CR} \left(\frac{\epsilon U_{\rm SN}}{\muu c^2}\right) \;
       \frac{X_j/A_j}{y_j^{\rm GCR}}
       \frac{(4\pi D^2 \Delta t)^{-1}}{\Phi} \\
 & = & 16 \ \frac{X_j/A_j}{y_j^{\rm GCR}}
       \left(\frac{D}{10 \ {\rm pc}}\right)^{-2}
\eeqar
where we have now assumed the production to be dominated by
the projectile species $j$.
The signal-to-background in this case is, of course,
much smaller than that for direct deposition.

Note also that at a distance of 40 pc the cosmogenic signal drops
below background.  But this is roughly the distance of the cutoff
for the direct supernova ejecta.  Thus it appears that
there is either a strong direct deposition signal for a
very nearby supernova, or perhaps a
feeble cosmic ray signal for one a little further, or no signal at all
for larger distances.

\subsection{Fission Background}

Fission of ambient, long-lived, heavy nuclei, notably \u238, leads to
significant production of some of the radioisotopes of interest.
Since \u238 dominates, we will simplify by only considering
this parent nucleus.
The radiological background will
be
$\Lambda_i^{\rm BG} = X_i/(A_i \muu)$, where
$X_i$ is the ambient mass fraction
in the ice or sediment of interest.
We thus need to compute the mass fraction of daughter species $i$
at a present time $t_0$, after the deposition time
$t_{\rm d}$; this is just given by the integrated U decay rate
times the branching ratio for spontaneous
fission into the species $i$.
This leads to a background of
\beq
\label{eq:fiss}
\Lambda_i^{\rm BG} = \frac{f_i}{A_i} \
   \frac{t_0 - t_{\rm d}}{\tau_{\rm U}^{\rm SF}} \
   \frac{X_{\rm U}}{\muu} \ \ .
\eeq
where $\tau_{\rm U}^{\rm SF} = 1.3 \times 10^{16}$\  yr
is the lifetime against spontaneous fission of \u238, and
$f_i$ is the fraction of fissions that produce $i$.

For a time since deposition $t_0 - t_{\rm d} = 100$ kyr (appropriate for
ice cores), eq.\ (\ref{eq:fiss}) gives a background level of
$\Lambda_i^{\rm BG} \sim f_i \ 1.5 \times 10^{2}$ atoms/g,
assuming that uranium has its average
(present) terrestrial abundance.  Note that
$f_i$ is fairly tightly distributed around
mass numbers $100$ and $140$.
While this background is tiny, the signal is as well;
indeed, fission can be an important background source for isotopes
near these peaks whose
cosmogenic background is small, notably \i129 and \sm146.
This is particularly true if one examines longer-lived
isotopes in deep sea sediments deposited on timescales of Myr ago.

\section{Isotope Candidates}
\label{sec:isotopes}

Having presented the various effects and backgrounds,
we turn to the possible isotope candidates,
both for probing the Geminga
event and for mass-extinction events
(specific signature predictions are presented in the next section).
Isotopes arise from either supernova explosions, or
cosmic ray production, and can be
further subdivided into short- and long-lived
radioactivities,.  They can thus be classified:
\begin{enumerate}
\item short-lived ($t_{1/2} < 10^7$ yr) SN products
\item long-lived ($t_{1/2} \ge 10^7$ yr) SN products
\item short-lived CR products
\end{enumerate}
as there are no long-lived CR products. Moreover,
as the cosmic-ray products are very few, the bulk of the
candidates are potential supernova products.  Of course,
a supernova origin would not be established equally easily
for all candidates of interest.
Note that while a few isotopes seem very likely to be
supernova products (e.g., \al26, \cl36, and \ni59),
for others this is less clear.
Indeed, turning the problem around,
detection of these isotopes could teach us about the source of
the different candidate nuclei.

Short-lived isotopes are good as
Geminga signatures or as extinct radioactivity;
it is clear that they are unable to provide signatures of
mass extinctions.
Good short-lived SN products are notably
\al26, \cl36 \fe{60}, and \ni59.
Indeed, there is now direct evidence for
\al26 in the ISM, observed via its 1.809 MeV $\gamma$-ray
emission line (Kn\"odlseder, Oberlack, Diehl, Chen, \& Gehrels 1996).
This emission is concentrated in the Galactic plane
and strongly suggests a supernova origin
for \al26 (see Prantzos \& Diehl 1996 for a review).
The only short-lived
cosmic-ray product produced in
a significant abundance is \be10,
(with a possible contribution to \al26 as well).

The long-lived isotopes can provide a long enough signal to give
evidence of a mass extinction.
We note first that while \k40 and \u238 might seem good candidates,
in fact they are not, precisely because their lifetimes are so long
($>1$ Gyr).  Their longevity has allowed a significant fraction of
their initial abundance to remain in ambient terrestrial matter.
For our purposes, this leads to the
same difficulties as the stable isotopes:
the ambient background overwhelms signal.
Thus, we wish to find isotopes sufficiently long-lived to
provide signatures of mass extinctions, but with lifetimes
that are still short compared to the age of the earth.
There are few of these.

Indeed, since the most interesting mass extinctions
occurred at epochs $\ga 10^7$ yr ago, there are only
two isotopes with lifetimes in this range, which
can be discussed individually.
\i129 ($\tau = 15$ Myr) is thought to be produced in the
r-process.  If this has its origin in supernovae, it
makes a good signature due to its small background (coming from cosmogenic
production via rare Xe targets); on the other hand, it is
too short-lived for the most ancient mass extinctions.  
\sm146 ($\tau = 146$ Myr) is produced via
the p-process 
This presumably has its site in
supernovae, though the protosolar abundance is
poorly reproduced by specific supernova models
(Prinzhoffer et al.\ 1989; Lambert 1992).
Alternatively, Woosley \& Howard (1990) have investigated
the possible orion of \sm146 via photodissociation in the
$\gamma$-process.
\pu244 ($\tau = 118$ Myr) comes from the r-process.
Thus it is not clear
that the long-lived nuclei are SN products. However,
even if if both are not made in
supernovae, they could appear
in the swept-up material due to their
ISM equilibrium abundance, if the nearby explosion occurs
in a dense ($n_{H} \ga 1 \ {\rm cm}^{-3}$) medium.

\section{Implications for the Geminga Supernova}

Anomalous \be10 abundances at $\sim 35$ and 60 kyr B.P. (before 
the present) were first reported for
antarctic ice cores taken in Vostok and Dome C
(Raisbeck et al.\ 1987). 
Recently, Raisbeck et al.\ (1992) have taken high-resolution 
data at Vostok, extending to 50 kyr B.P{}.  They reconfirm
the presence of the 35 kyr peak, which persists and is even 
amplified
when correcting for variations in the precipitation 
rate.\footnote{Raisbeck et al.\ (1992) also discuss the possible 
relationship between
the \be10 enhancement and anomalies in the \c14/\c12 ratio.
For epochs prior to $\sim 10$ kyr B.P., 
there is a discrepancy between \c14 and U-Th dating methods;
assuming the latter to be accurate, the \c14/\c12 ratio
shows an unexplained rise reaching to the end of the 
available data ($\sim 23$ kyr B.P.).  Raisbeck et al.\ 
note that there is qualitative agreement between the
\be10 and \c14 behaviors.  However, they argue that
apparent variations in the \c14/\be10 ratio suggest
that the cosmogenic production had a different energy 
spectrum than at present, perhaps due to, e.g., 
transients in the development of cosmic ray shock acceleration.}
Other groups have now confirmed the \be10 enhancements.
Beer et al.\ (1992) report positive detections 
of the 35 kyr peak
elsewhere in 
the antarctic (Byrd station), as well as in Camp Century, Greenland 
ice cores (although they cannot confirm the 60 kyr peak in either sample).
A \be10 peak has also been seen at $\sim 35$ kyr 
in deep sea sediments off
the Gulf of California (McHargue, Damon, \& Donahue 1995) and in the
Mediterranean (Cini Castagnoli et al.\ 1995).
That the enhancement has been seen in these diverse locations
and media strongly suggests that the effect was indeed a global
one, as we would predict.  

On the other hand, it has been suggested
(Mazaud, Laj, \& Bender 1994) that the \be10 spikes may derive from
variations in the geomagnetic field.  To obtain a good correlation,
these authors use a ice deposition rate history different from the
most recent calculations.  To be sure, there remains some correlation
with geomagnetic intensity, which may explain part of the \be10
enhancements\footnote{Although McHargue et al.\ (1993, 1995) suggest
that even this might be attributed to extraterrestrial causes.}.

For the most part, though,
discussion of the anomalous \be10 measurements
has focussed on direct passage of the shock wave past the Earth
(Raisbeck et al.\ 1987;
Sonett, Morfill, \& Jokipii 1987;
Ammosov et al.\ 1991;
Sonett 1992;
Ramadurai 1993;
Lal \& Jull 1992).
This work has concluded that the Vostok data
may indicate a supernova explosion occurred
at distances of $\la 100$ pc, and perhaps
even shows something of the detailed shock structure.
The ``double-bump'' structure of the Vostok \be10 anomaly could
conceivably be due to the shock wave bouncing back from the
boundary of a previously-cleared low-density bubble
(Frisch 1994) in the ISM.
Further, 
Ramadurai (1993) has suggested
that the supernova
causing the \be10 might be Geminga itself.
Indeed, recent {Hubble Space Telescope} observations 
(Caraveo, Bignami, Mignani, \& Taff 1996)
have measured Geminga's parallax and so determined that
it is quite close, at a distance
of $157 \ {}^{+57}_{-34}$ pc.\footnote{This distance
implies transverse velocity of
122 km/s, while the radial velocity remains unknown.  
The direction of transverse motion is consistent with
the suggestion (Smith, Cunha, \& Plez 1994)
that Geminga originated in Orion.  However, for this to be the
case, Geminga would have to have a very large radial velocity,
making it one of the fastest pulsars known.  In any case, 
the origin site is tied to the age estimate, and so does not
resolve this issue.}

However, there may be problems reconciling the Geminga event dating
implied by the \be10 anomalies with
dating estimates from pulsar spin-down arguments
(Halpern \& Holt 1992; Bignami \& Caraveo 1992; 
Bertsch et al.\ 1992).
The former gives something like 60--100 kyr,
while the latter give something more like 300 kyr.
One should bear in mind, though, that the spin-down times
give an upper bound to the time since the explosion, as
neutron starquakes can lead to very rapid mass redistribution
and slowing of angular velocity, know as ``glitches.''
Indeed, Alpar, \"Ogelman, \& Shaham (1993) have argued that 
Geminga is indeed a glitching pulsar.
If there were a number of such events, then Geminga might be
more recent and the age estimates could be brought into 
agreement.  
Moreover, as noted in \S \ref{sec:cgen}, the signals will be
delayed by the propagation time, which will be significant for
the direct debris, and for any cosmic ray component entangled in
the shock.  In this case, a time delay of order 300 kyr would
indicate a distance $\sim 30$ pc.

Despite the possible difficulties in reconciling the age determination,
it is in any case interesting to consider
eq.\ (\ref{eq:cgsb}) in the light of the Vostok
ice-core data.
Let us
assume the Vostok \be10 peaks are due to a supernova.
Then one may ask:  what distance does this imply?
In the data, the signal-to-background ratios for the
peaks fall within the generous
range of $1 \le \Lambda_{\rm peak}/\Lambda_{\rm BG} \le 4$.
Interpreting the peaks as signal, eq.\ (\ref{eq:cosgen}) implies that
$20 \ {\rm pc} \la D \la 40 \ {\rm pc}$.  This suggests that if the
Vostok peaks came from a supernova, it was quite close and indeed
may have been a near miss.

With this result in hand, we have
collected the predictions for all isotopic
signatures and backgrounds in table 1.
To fix ideas, we have calculated the entries in the table
for a supernova at $D= 20$ pc, and the specific
abundances $\Lambda_i$ are for ice core sedimentation
rates.
The scalings with distance
for each component have been noted both in the text and in figure 2.

The signals computed in table 1 come from several
recent nucleosynthesis calculations.  The supernova yields
of \al26, \cl36, \ca41, \mn53, \ni59, and \fe{60} are
taken from Woosley \& Weaver (1996), for their
$20 \msol$ model S20A{}.
R- and p-process yields of \i129, \sm146, and \pu244
are taken from Cameron, Thielemann, \& Cowan (1993).
The table takes the optimistic view that
these long-lived isotopes all have their
origin in supernovae.  Also, the swept-up component assumes
(optimistically) a dense ($n_{\rm ISM} = 1$)
local ISM.

Note also that some signatures are best observed not as
an absolute abundance in atoms/g, but in terms of
an isotopic fraction, e.g., \cl36/Cl.  For these last cases, the
expected signature can be deduced from the known background
isotopic fraction, and the signal-to-background ratio as deduced
from the table.

If the \be10 signal does have its origins in the Geminga blast,
then table 1 indicate that
several other isotopes should be much more abundant in the ice cores.
Let us take \al26 as an example.
So long as Geminga occurred within $D \la D_{\rm max}$, then
we expect $\al26/\be10 \simeq 4$.
Detection of \al26 spikes at the same strata as those of \be10
would lend strong support to the notion of a nearby supernova origin
for the Vostok \be10 signal.  Further,
since the \be10 component arises from enhanced cosmogenic production,
where the \al26 component
is dominated by direct deposition,
the detection of the latter would also confirm that both
mechanisms indeed happen and are
important.

In this regard,
it is interesting that Cini Castagnoli et al.\ (1992)
indicate that they performed two \al26 measurements on their
Mediterranean core at the peak regions.  They do not give
a quantitative result, but cite this measurement as evidence against
a contribution from cosmic dust.  The implication is that there was
not a large \al26 signal.  While this certainly does not
strengthen the Geminga
hypothesis,  it cannot rule it out.  For example,
\al26 might not be a supernova product (though
interstellar 
$\gamma$-ray line observations argue against this), or
the direct ejecta may have been excluded from the inner solar system.
Further and more quantitative data would be
very helpful in resolving this
question.  For example, the lack of signal in
other relatively abundant cosmogenic nuclides,
such as \cl36, would be strong evidence against
the Geminga interpretation.

Of course, aside from \al26, the other
isotopes we have listed are potentially detectable as well.
Indeed, \mn53 and \ni59 could even be at
somewhat higher levels.  Note also the variety
of likely candidates; this helps insure that the possibility of
a signal is not overly tied to uncertainties about
the supernova origin of a particular radionuclide.

We hope that the promising outlook embodied in
table 1 will prompt a search for these isotopes
in the ice cores.  Even a null signal would be an important
indication
that the \be10 peaks are {\it not} due to a supernova.
Also, it is important to note that in ocean sediments,
the low level of precipitation makes the signal-to-background
ratio larger by a factor of $\sim 10^3$.  Thus these
could provide even clearer evidence of a nearby supernova.

\section{Conclusions}

We have considered in this paper various origins for geological
isotope anomalies as possible signatures of nearby supernova explosions,
including the supernova ejecta themselves,
material swept up from the ISM, and isotopes produced by
cosmic-ray collisions in the atmosphere.
We have explored the prospects for searches in ice cores.
These could be useful in understanding the origin of the
global \be10 anomalies and possibly finding a trace of the
Geminga explosion.
We have also considered searches in deep-ocean sediments, which
could provide evidence for any supernova explosion near enough
to have affected the biosphere and possibly caused a mass extinction.
We have explored the possibilities of searches for live and extinct
radioactivities, and for low-level trace abundances.

The best prospects seem to be offered by searches for trace
amounts of supernova ejecta. This may be considerably stronger than the
background induced by conventional cosmic rays. The atmospheric
production of spallation isotopes by cosmic rays from a nearby
supernova explosion may be observable if the supernova was
sufficiently close, namely within about $40$ pc.

Table 1 lists the shorter-lived radioisotope candidates that
are of particular interest for searches in ice cores,
which may extend back to about $300$ kyr B.P{}. The isotopes
\al26, \ca41, \ni59 and \fe{60} may be the most promising signatures
of a nearby supernova such as Geminga during this period.
It would be very interesting to look for a
correlation with the Vostok \be10 anomalies, to test the
hypothesis that these could be due to the Geminga or another
nearby supernova. We re-emphasize that this identification does
not seem exceedingly likely, given the usual estimates of the
age and distance of the Geminga remnant 
(Halpern \& Holt 1992; Bignami \& Caraveo 1992; 
Bertsch et al.\ 1992), but
cannot be excluded and should be explored.

Also included in table 1
are longer-lived radioisotopes that could be
of interest for searches in deep-ocean sediments, which may
extend back to several hundred Myr B.P{}.
\i129 is produced by cosmic rays in the atmosphere, and has
a small background.
\sm146 could be produced in supernovae via the p-process.
Although the origin
of the r-process (and thus \pu244) is unclear, it should be present in the ISM,
and their detection could tell us something about the source of
r-process nuclei.

The abundances of all isotopes depend strongly
on the distance of any supernova explosion, in different
ways for different production mechanisms. Thus a deep-ocean
sediment search may be able to tell us whether an explosion
could have occurred sufficiently nearby (less than about $10$ pc)
to have affected the biosphere, or whether there might have been
a ``near miss.'' However, we re-emphasize that our estimates of the
possible abundances do not take into account fractionation, which
could be important for deep-ocean sediments.
In addition, the low precipitation rate for ocean sediments makes signal
more pronounced than in
ice cores by a factor $\sim 10^3$.  So if can indeed anomalies
are found to be observable in ice
cores, they should stand out clearly indeed in ocean sediments (so long
as the isotopes are sufficiently long-lived).

Any radioisotope signal above the background from conventional
sources would provide a unique tool, not only to learn about a
possible mechanism for a mass extinction, but possibly also about
supernovae themselves and the various processes that synthesize
different elements in the cosmos.  We are encouraged that the
prospects are good for the detection of a supernova signal over
background, and we encourage experimental searches for such signatures.

\acknowledgements
We are pleased to acknowledge conversations
with Walter Alvarez, Robert Moch\-ko\-vitch, George Reid, and
Jim Truran.
This material is based upon work supported by the North Atlantic
Treaty Organization under a Grant awarded in 1994.
D.N.S. is supported by the NSF, by NASA and by the DOE at the
University of Chicago and by the DOE and by NASA through grant NAG5-2788 at
Fermilab.

\newpage

\centerline{Figure Captions }

\bigskip

\figcaption[]
{\label{fig:mvd}
Deposited mass as a function of distance $D$ from the supernova.
The total mass deposited is shown, as well as
the component due to direct deposition and to cosmogenic production.
Note the increase of material above about 7 pc, which continues
until the cutoff at $\sim 45$ pc.  Note that the deposited mass will
scale directly with the ISM density, while the
cutoff will scale with the solar wind pressure at earth and inversely
with the blast duration.  Although the cosmogenic
contribution is negligible when there is a direct component, it
is the only source above the cutoff.}

\figcaption[]
{\label{fig:obs}
Expected number of radioisotopes per
unit mass of sediment, $\Lambda$.
Cosmic ray backgrounds
and detection sensitivity are indicated.
The direct deposition yields are for \al26, cosmogenic yields
for \be10.  These can be scaled using Table 1 to give
the dependences for other isotopes.}

\newpage

\begin{table}[htb]
\label{tab:sigs}
\begin{center}
\caption{Ice core signatures for a supernova at 20 pc 
(in atoms/g) }
\begin{tabular}{|ccccc|ccc|}
\hline \hline
Isotope & SN ejecta & Swept & Cosgen.\ & {\sc tot.\ signal}
& Cosgen.\ bgd.\ & Rad.\ bgd.\ & {\sc tot.\ bgd.\ } \\
\hline
\be10 & --- & --- & \scn{1.9}{6} & \scn{1.9}{6}
   & \scn{2.2}{6} & --- & \scn{2.2}{6} \\
\al26  & \scn{8.4}{6} & \scn{9.3}{4} & \scn{3.1}{3} & \scn{8.4}{6}
   & \scn{3.5}{3} & --- & \scn{3.5}{3}\\
\cl36  & \scn{4.8}{6} & \scn{2.2}{4} & \scn{6.6}{4} & \scn{4.9}{6}
   & \scn{7.5}{4} & --- & \scn{7.5}{4} \\
\ca41 & \scn{1.5}{6} & \scn{6.7}{3} & 1.4 & \scn{1.5}{6} 
   & 1.6 & --- & 1.6 \\
\mn53 & \scn{2.3}{7} &\scn{1.3}{6} & 0.7 & \scn{2.4}{7}
   & 0.79 & --- & 0.79 \\
\ni59 & \scn{1.0}{7} & \scn{1.2}{4} & --- & \scn{1.0}{7}
   & \scn{1.6}{-3} & --- & \scn{1.6}{-3} \\
\fe{60} & \scn{1.2}{6} & \scn{5.4}{3} & 1.4 & \scn{1.2}{6}
   & 1.6 & --- & 1.6 \\
\i129 & \scn{6.9}{3} & \scn{1.70}{3} & 1.4 & \scn{8.6}{3}
   & 1.6 & 1.1 & 2.7 \\
\sm146 & 0.32 & 0.50 & --- & 0.82
   & --- & 6.7 & 6.7 \\
\pu244 & 69 & 86 & --- & \scn{1.6}{2}
   & --- & --- & --- \\
\hline \hline
\end{tabular}
\end{center}
\end{table}


\begin{thebibliography}{}

\bibitem{aaam} Alvarez, L., Alvarez, W., Asaro, F., \& Michel, H. 1980,
Science, 208, 1095

\bibitem{aos} Alpar, M.A., \"Ogelman, H., \& Shaham, J. 1993, A\&A, 273, L35

\bibitem{amm} Ammosov, A.E., et al., Izv. Akad. Nauk. SSSR, Ser. Fiz. 1991
55, 10

\bibitem{abkw}  Arnett, W.D., Bahcall, J., Kirshner, R.
\& Woosley, S. 1989, ARAA, 27, 629

\bibitem{beer} Beer, J., et al., in The Last Deglaciation:
Absolute and Radiocarbon Chronologies (ed.\ E. Bard \& W.S. Broecker)
Natio ASI Series, 12 (1992: Hiedelberg, Springer-Verlag), 141

\bibitem{bry} Beer, J., Raisbeck, G.M., \& Yiou, F., in
The Sun in Time, ed.\ C.P. Sonett, M.S. Giampapa, \& M.S. Mathews
(1991: Tucson, Univ.\ of Arizona Press), 343

\bibitem{bertsch} Bertsch, D.L., et al.\ 1992, Nature, 357, 306

\bibitem{bc} Bignami, G.F., \& Caraveo, P.A. 1992, Nature, 357, 287

\bibitem{ctc} Cameron, A.G.W., Thielemann, F.-K., \& Cowan, J.J.
1993, Phys.\ Rept., 227, 283

\bibitem{cbmt} Caraveo, P.A, Bignami, G.F., Mignani, R. \& Taff, L.G.
1996, ApJ, 461, L91

\bibitem{cl} Chevalier, R.A., \& Liang, E.P. 1989, ApJ, 344, 332

\bibitem{med} Cini Castagnoli, G., et al.\ 1995,
Geophys. Res. Lett., 22, 707

\bibitem{cb} Crutzen, P.J., \& Br\"uhl, C. 1995,
Max-Planck-Institut f\"ur Chemie, Mainz preprint

\bibitem{es} Ellis, J. \& Schramm, D.N. 1995,
Proc. Nat. Acad. Sci., 92. 235, (1995)

\bibitem{frisch} Frisch, P.C. 1994, Science, 256, 1423

\bibitem{gc}  Gehrels, N., and Chen, W. 1993, Nature, 361, 706

\bibitem{hh} Halpern, J.P., \& Holt, S.S. 1992, Nature, 357, 222

\bibitem{kodcg} Kn\"odlseder, J., Oberlack, U., Diehl, R.,
Chen, W., \& Gehrels, N. 1996, A\&A, in press (astro-ph/9604057)

\bibitem{lj} Lal, D., \& Jull, A.J.T. 1992, Radiocarbon, 43, 227

\bibitem{lambert} Lambert, D.L., 1992, Astron.\ Astrophys.\ Rev., 3, 201

\bibitem{lpw} Lee, T., Papanastassiou, D.A.,\& Wasserburg, G.J.
1977, ApJ, 211, L107

\bibitem{mlb} Mazaud, A. Laj, C, \& Bender, M. 1994,
Geophys. Res. Lett., 21, 337

\bibitem{mccray} McCray, R. 1993, ARAA, 31, 175

\bibitem{mdd93} McHargue, L.R., Damon, P.E., \& Donahue, D.J. 1993,
in proceedings of the XXIII ICRC Conference, 3, 854

\bibitem{mdd} McHargue, L.R., Damon, P.E., \& Donahue, D.J. 1995
Geophys. Res. Lett., 22, 659

\bibitem{mckee} McKee, C.F., in IAU Colloquium 101, The Interaction of
Supernova Remnants with the Interstellar Medium, ed. T. Landrecker
and R. Rogers (1988: Cambridge, Cambridge Univ.\ Press), 205

\bibitem{meyer} Meyer, B. et al.\ 1992, ApJ, 399, 656

\bibitem{ms} Meyer, B.S., \& Schramm, D.N. 1986, ApJ, 311, 406

\bibitem{odas} O'Brien, K., de la Zerda Lerner, A., Shea, M.A., \& Smart,
D.F. in The Sun in Time, ed.\ C.P. Sonett, M.S. Giampapa, \& M.S. Mathews
(1991: Tucson, Univ.\ of Arizona Press), 317

\bibitem{parker} Parker, E.N. 1958, Phys.\ Rev., 110, 1445

\bibitem{perko} Perko, J.S. 1987, A\&A, 184, 119

\bibitem{pd} Prantzos, N., \& Diehl, R. 1996, Phys.\ Rep., 267, 1

\bibitem{prinz} Prinzhoffer et al., 1989 ApJ, 344, L81

\bibitem{rais} Raisbeck, G.M., et al.\ 1987, Nature, 326, 273

\bibitem{rais92} Raisbeck, G.M., et al., in The Last Deglaciation:
Absolute and Radiocarbon Chronologies (ed.\ E. Bard \& W.S. Broecker)
Natio ASI Series, 12 (1992: Hiedelberg, Springer-Verlag), 127

\bibitem{rama} Ramadurai, S. 1993, Bull. Astr. Soc. India, 21, 391

\bibitem{rmc} Reid, G.C., McAfee, J.R., Crutzen, P.J. 1978, Nature, 257, 489

\bibitem{rud} Ruderman, M.A. 1975, Science, 184, 1079

\bibitem{scp} Smith, V.V., Cunha, K., \& Plez, B. 1994, A\&A, 281, L4

\bibitem{son} Sonett, C.P. 1992, Radiocarbon, 34, 2

\bibitem{smj} Sonett, C.P., Morfill, G.E., \& Jokipii, J.R. 1987,
Nature, 330, 458

\bibitem{spitz} Spitzer, L. 1978, Physical Processes in the Interstellar Medium
(Wiley:  New York), 255

\bibitem{van} van den Bergh, S. 1994, Pub.\ Astron.\ Soc.\ Pacif., 106, 689

\bibitem{wh} Woosley, S.E., \& Howard, W.M. 1990, ApJ, 354, L21

\bibitem{ww} Woosley, S.E., \& Weaver, T.A. 1996, ApJS, 101, 81

\end{thebibliography}
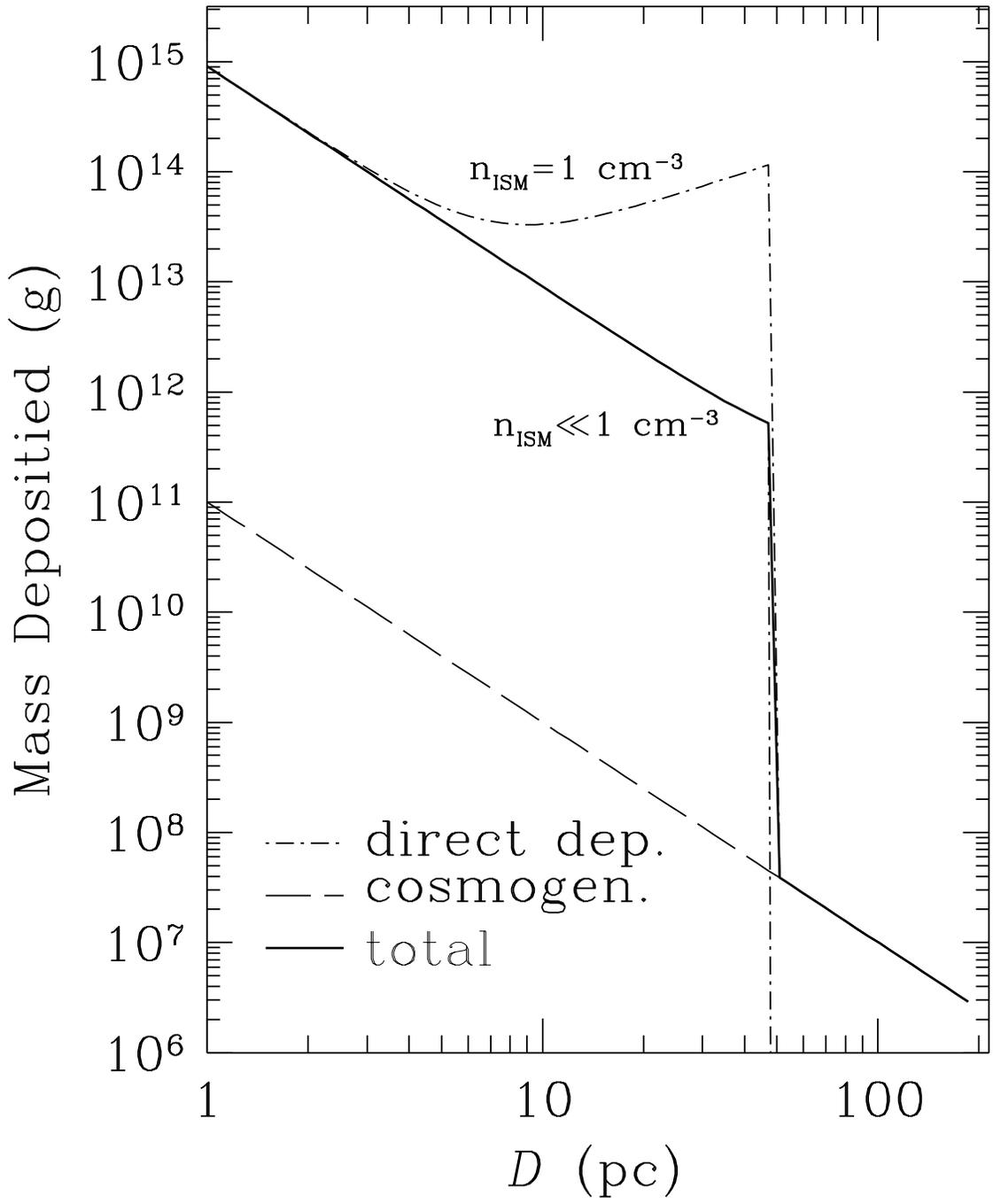
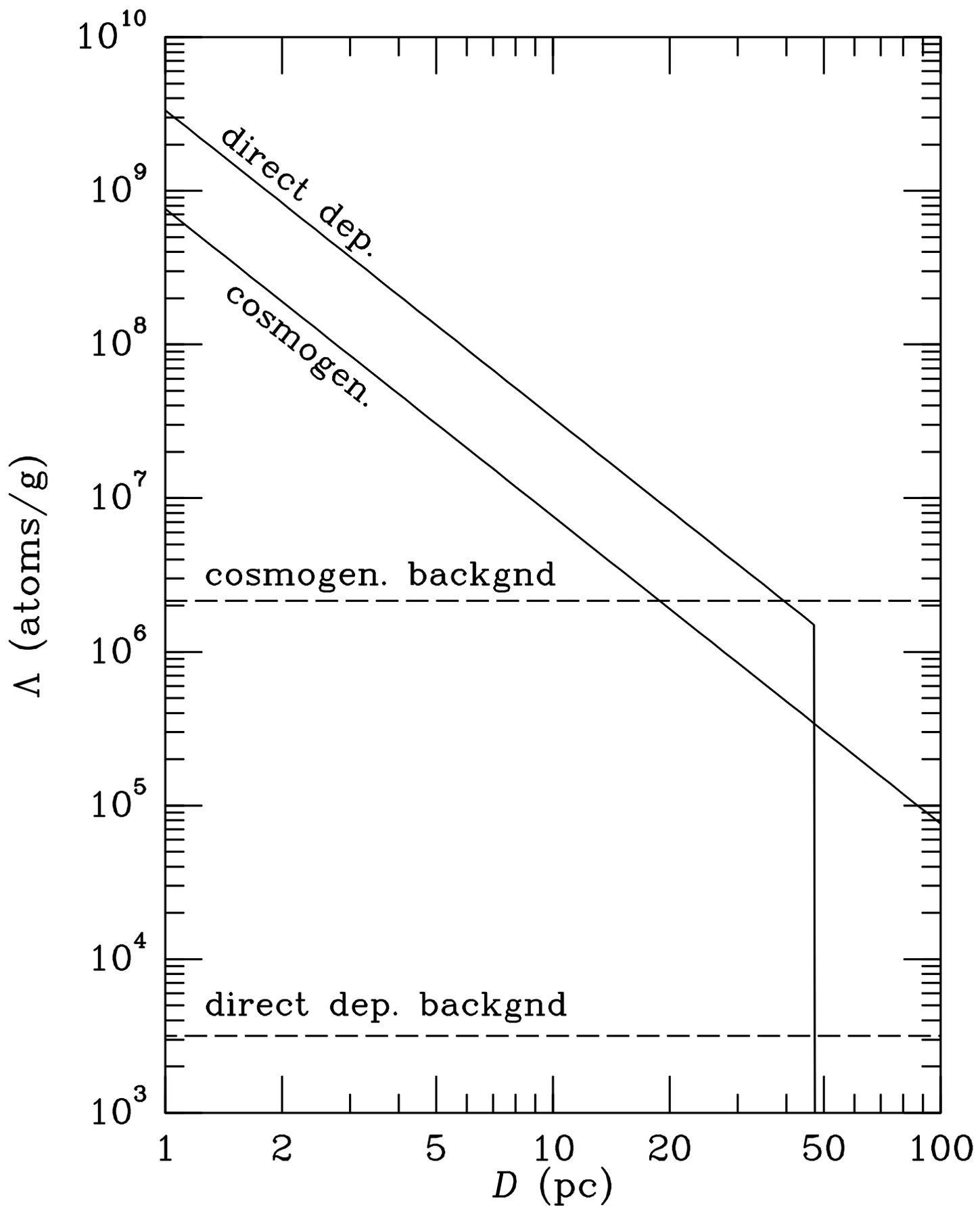
\end{document}